\documentclass[runningheads]{llncs}
\usepackage[T1]{fontenc}
\usepackage{graphicx}
\usepackage{multirow}
\usepackage{amsmath}
\usepackage{amssymb}
\usepackage{booktabs}
\usepackage[dvipsnames]{xcolor}
\usepackage[colorlinks=true, linkcolor=blue, urlcolor=blue, citecolor=blue, anchorcolor=blue]{hyperref}

\begin{document}
\title{Weakly Supervised Learning of Cortical Surface Reconstruction from Segmentations}

\titlerunning{Weakly Supervised Cortical Surface Reconstruction}
%
\author{Qiang Ma\inst{1}\and
Liu Li\inst{1}\and
Emma C. Robinson\inst{2}\and
Bernhard Kainz\inst{1,2,3}\and\\
Daniel Rueckert\inst{1,4}
}


\authorrunning{Q. Ma et al.}

\institute{Department of Computing, Imperial College London, UK \\
\email{q.ma20@imperial.ac.uk}
\and
School of Biomedical Engineering and Imaging Sciences,\\King's College London, UK
\and
FAU Erlangen–N\"urnberg, Germany
\and
Klinikum rechts der Isar, Technical University of Munich, Germany
}

\maketitle              
\begin{abstract}

Existing learning-based cortical surface reconstruction approaches heavily rely on the supervision of pseudo ground truth (pGT) cortical surfaces for training. Such pGT surfaces are generated by traditional neuroimage processing pipelines, which are time consuming and difficult to generalize well to low-resolution brain MRI, \emph{e.g.}, from fetuses and neonates. In this work, we present CoSeg, a learning-based cortical surface reconstruction framework weakly supervised by brain segmentations without the need for pGT surfaces. CoSeg introduces temporal attention networks to learn time-varying velocity fields from brain MRI for diffeomorphic surface deformations, which fit an initial surface to target cortical surfaces within only 0.11 seconds for each brain hemisphere. A weakly supervised loss is designed to reconstruct pial surfaces by inflating the white surface along the normal direction towards the boundary of the cortical gray matter segmentation. This alleviates partial volume effects and encourages the pial surface to deform into deep and challenging cortical sulci. We evaluate CoSeg on 1,113 adult brain MRI at 1mm and 2mm resolution. CoSeg achieves superior geometric and morphological accuracy compared to existing learning-based approaches. We also verify that CoSeg can extract high-quality cortical surfaces from fetal brain MRI on which traditional pipelines fail to produce acceptable results.

\end{abstract}

\section{Introduction}

Cortical surfaces, \emph{i.e.}, the inner/white and outer/pial surfaces of the cerebral cortex, play a crucial role in visualizing the anatomical structure as well as quantitative characterizing the morphology of the cortex. Traditional neuroimage processing pipelines~\cite{dai2013ibeat,fischl2012freesurfer,glasser2013hcp,makropoulos2018dhcp} achieved great success for cortical surface reconstruction from adult or neonatal brain MRI. However, these pipelines normally incorporate multiple processing steps and require more than 6 hours for a single subject. The limited accuracy of prior processing steps will also cause subsequent corruptions in cortical surfaces. Moreover, traditional pipelines contain a large amount of carefully tuned parameters, which make them difficult to generalize across data domains in different age groups or acquisition protocols.

To boost the accuracy and efficiency of cortical surface reconstruction, recent studies~\cite{cruz2021deepcsr,gopinath2023clinical,henschel2020fastsurfer,wang2023ibeat} leverage deep learning-based approaches to predict implicit surface representations, \emph{i.e.}, segmentations and signed distance functions (SDF). Explicit 3D meshes are extracted from the implicit surfaces by Marching Cubes (MC) algorithm~\cite{lorensen1998marching}. Topology correction algorithms~\cite{bazin2005topology,segonne2007topology} are used to detect and repair topological errors such that the extracted surfaces are homeomorphic to a 2-sphere. However, this is a time-consuming process based on iterative refinement, which is hard to accelerate efficiently through engineering efforts. 
In addition, although existing approaches are capable of producing accurate segmentations~\cite{billot2023synthseg,henschel2020fastsurfer,roy2019quicknat,uus2023bounti,wang2023ibeat} or SDFs~\cite{cruz2021deepcsr,gopinath2023clinical}, these volumetric representations intrinsically face the challenges of partial volume effects, leading to difficulties for the MC algorithm in accurately capturing the cortical folding of extracted surfaces. This issue is especially pronounced within the cortical sulci, in particular for low-resolution brain MRI such as fetal and neonatal data.

Instead of learning implicit surfaces, latest works~\cite{bongratz2022vox2cortex,bongratz2024v2c-flow,chen2023surfflow,hoopes2022topofit,lebrat2021corticalflow,ma2023cotan,ma2022cortexode,ma2021pialnn,santa2022cfpp,zheng2024coupled} focus on learning explicit surface deformations end-to-end from brain MRI. Such approaches only require a few seconds to deform an initial mesh to target cortical surfaces. However, one major limitation is that they heavily rely on the supervision of pseudo ground truth (pGT) cortical surfaces generated by traditional pipelines~\cite{dai2013ibeat,fischl2012freesurfer,glasser2013hcp,makropoulos2018dhcp} such as FreeSurfer~\cite{fischl2012freesurfer}. Their long processing time makes it expensive to collect a large dataset for training. Also, errors in pGT surfaces inevitably form an upper bound for the achievable accuracy. 
Moreover, traditional pipelines may fail to extract pGT surfaces from fetal or clinical MRI sequences, which are inherently limited in scan time and thus lower image resolution.

\textit{Contributions:} We present CoSeg, a weakly supervised cortical surface reconstruction framework without the need for pGT surfaces during the training phase. CoSeg learns cortical surfaces explicitly with the weak supervision of cortical ribbon segmentations, which are more accessible than pGT cortical surfaces particularly for low-resolution images~\cite{billot2023synthseg,uus2023bounti}. We introduce temporal attention networks (TA-Net)~\cite{ma2023cotan} to learn diffeomorphic surface deformations from brain MRI. To address partial volume effects on the pial surfaces, we propose a novel weakly supervised loss consisting of a boundary loss, which inflates the white surface towards the boundary of the cortical gray matter (cGM) segmentation, as well as an inflation loss, which constraints the inflation to follow the normal direction. We evaluate CoSeg on the HCP young adult dataset~\cite{van2013hcp} and provide qualitative evaluation results for the dHCP fetal dataset~\cite{price2019fetal}. The code for CoSeg are released publicly at \url{https://github.com/m-qiang/CoSeg}.

\section{Method}

\begin{figure}[ht]
\centering
\includegraphics[width=1.00\textwidth]{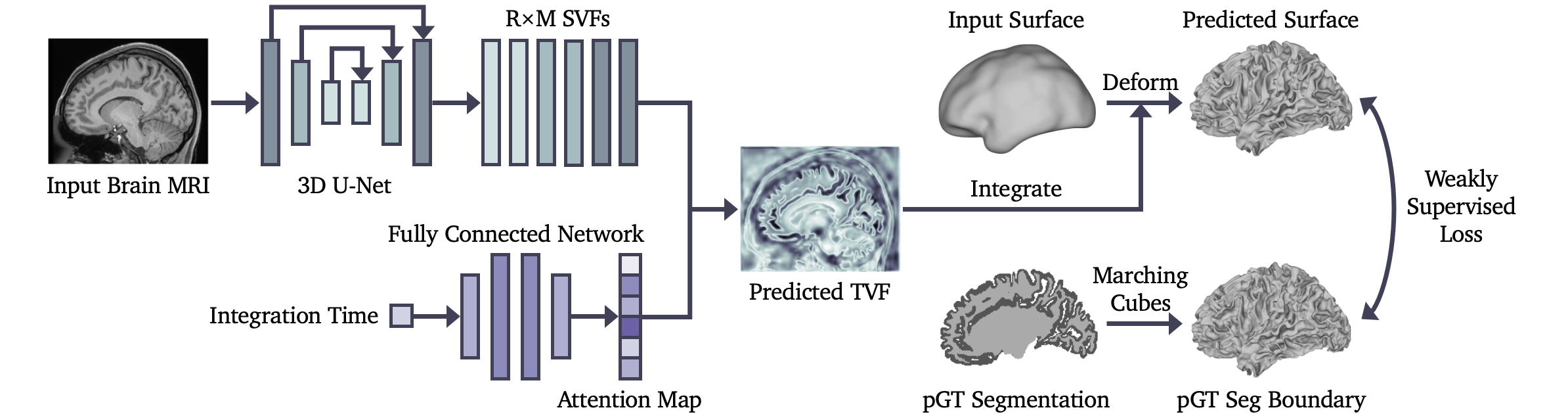}\label{fig:architecture}
\caption{The architecture of the TA-Net. TA-Net predicts a TVF by learning a weighted sum of multiple SVFs. The TVF is integrated to produce diffeomorphic surface deformations which fit an initial input surface to the predicted cortical surface.}
\end{figure}

\noindent\textbf{Temporal Attention Networks.} Given an initial surface $\mathcal{S}_0\subset\mathbb{R}^3$ with points $x_0\in\mathcal{S}_0$, a diffeomorphic surface deformation can be defined via an ordinary differential equation (ODE):
\begin{equation}\label{eq:ode}
\dot{x}(t)=u(x(t),t),~x(0)=x_0,~t\in[0,T],
\end{equation}
where $x(t)$ is the deformed surface point, and $u(x,t)\in\mathbb{R}^3$ is a learnable time-varying velocity field (TVF). By integrating the flow ODE (\ref{eq:ode}), the initial surface $\mathcal{S}_0$ is deformed to a target surface $\mathcal{S}_T$ with the points $x_T=x(T)$.

Since our CoSeg framework is model agnostic, we introduce a temporal attention network (TA-Net) based on CoTAN~\cite{ma2023cotan} to learn the TVF $u(x,t)$. While CoTAN is conditioned on the ages of the subjects, TA-Net is only conditioned on the integration time $t$ such that it can be extended to adult data. As shown in Fig.~\ref{fig:architecture}, given an input brain MRI image, TA-Net uses a 3D U-Net~\cite{ronneberger2015unet} to learn $M=2$ stationary velocity fields (SVF) for each of $R=3$ resolution levels. Thus, there are total $R$$\times$$M$ SVFs denoted by $\mathbf{u}(x)\in\mathbb{R}^{RM\times3}$. Given an input integration time $t\in[0,T]$, a channel-wise time-varying attention map $\mathbf{p}(t)\in\mathbb{R}^{RM}$, which satisfies $\sum\nolimits_{i=1}^{RM}\mathbf{p}_i(t)=1$, is predicted by a fully connected network to measure the importance of all $R$$\times$$M$ SVFs. By computing the weighted sum over all SVFs, the TVF is represented as $u(x,t)=\mathbf{u}(x)^{\top}\mathbf{p}(t)$.

We discretize a surface $\mathcal{S}$ by a 3D mesh $\mathcal{M}=(\mathcal{V},\mathcal{F},\mathcal{E})$, where $\mathcal{V},\mathcal{F},\mathcal{E}$ are the sets of vertices, faces, and edges respectively. Then, we integrate the ODE (\ref{eq:ode}) with the forward Euler method. For each integration step $k=0,...,K-1$, the vertices can be updated by $v_{k+1}=v_k+hu(v_k,hk)$, where $h=T/K$ is the step size and $v_0\in\mathcal{V}_0$ is the vertex of an input surface mesh $\mathcal{M}_0$. By integration, the TA-Net deforms the input mesh $\mathcal{M}_0$ to a predicted cortical surface mesh $\hat{\mathcal{M}}$ with vertices $\hat{v}=v_K$. The TA-Net can be trained by minimizing a weakly supervised loss between the predicted mesh $\hat{\mathcal{M}}$ and the pGT segmentation boundary.
\newline

\noindent\textbf{Initial Surface Generation.} We affinely align all brain MRI and pGT cortical ribbon segmentations to a standard space such as MNI-152. A fixed initial surface for all subjects is extracted from a template ribbon segmentation, which is obtained from either an atlas or the group average of all pGT segmentations in the training set. 
Inspired by~\cite{ma2022cortexode,santa2022cfpp}, we use a distance transform algorithm~\cite{breu1995distance} to convert the template white matter (WM) segmentation to a SDF, where the interior/exterior of the segmentation has negative/positive value. The SDF is smoothed by a Gaussian filter with standard deviation $\sigma$=6.0. An initial mesh $\mathcal{M}_0$ is extracted by MC~\cite{lorensen1998marching} from the SDF at level $l$=1.5. Such an initial mesh has genus-0 topology since the implicit surface has been sufficiently smoothed and inflated. We iteratively apply Laplacian smoothing and isotropic explicit remeshing~\cite{cignoni2008meshlab} to the initial mesh $\mathcal{M}_0$ until it has the desired number of vertices.
\newline

\noindent\textbf{White Surface Reconstruction.} CoSeg learns cortical surfaces weakly supervised by cortical ribbon segmentations, which contain the label maps of WM and cGM. The WM label includes subcortical structures such as deep gray matter and ventricle. The pGT segmentations can be obtained from either advanced learning-based approaches~\cite{billot2023synthseg,dai2013ibeat,henschel2020fastsurfer,roy2019quicknat,uus2023bounti} or manual annotations.

To learn the white surface, we minimize the commonly used bidirectional Chamfer (bi-Chamfer) distance $\mathcal{L}_{\mathrm{chamfer}}$~\cite{bongratz2022vox2cortex,fan2017chamfer,lebrat2021corticalflow,ma2023cotan,ma2022cortexode,santa2022cfpp,wickramasinghe2020voxel2mesh} between the predicted white surface mesh $\hat{\mathcal{M}}$ and the boundary of the pGT WM segmentation. The boundary is represented as a 3D surface mesh $\mathcal{M}_*$ extracted from the WM segmentation by MC~\cite{lorensen1998marching}. Taubin smoothing~\cite{taubin1995curve} is applied to alleviate the grid artifacts of the extracted mesh. Then, the bi-Chamfer loss~\cite{fan2017chamfer} is defined as:
\begin{equation}\label{eq:bi-chamfer-loss}
\mathcal{L}_{\mathrm{chamfer}}(\hat{\mathcal{M}}, \mathcal{M}_*)=\frac{1}{|\hat{\mathcal{V}}|}
\sum\nolimits_{\hat{v}\in\hat{\mathcal{V}}}
\min_{v_*\in\mathcal{V}_*}\|\hat{v}-v_*\|^2+
\frac{1}{|\mathcal{V}_*|}
\sum\nolimits_{v_*\in\mathcal{V}_*}
\min_{\hat{v}\in\hat{\mathcal{V}}}\|v_*-\hat{v}\|^2,
\end{equation}
where $\hat{v}$ and $v_*$ are the vertices of the predicted mesh $\hat{\mathcal{M}}$ and the pGT WM segmentation boundary $\mathcal{M}_*$ respectively. We add the edge length loss $\mathcal{L}_{\mathrm{edge}}$ and normal consistency loss $\mathcal{L}_{\mathrm{nc}}$~\cite{bongratz2022vox2cortex,wickramasinghe2020voxel2mesh} to enforce surface smoothness. The final loss is defined as $\mathcal{L}_{\mathrm{white}}=\mathcal{L}_{\mathrm{chamfer}}+w_{\mathrm{edge}}\mathcal{L}_{\mathrm{edge}}+w_{\mathrm{nc}}\mathcal{L}_{\mathrm{nc}}$ with weights $w_{\mathrm{edge}}$ and $w_{\mathrm{nc}}$.
\newline

\begin{figure}[t]
\centering 
\includegraphics[width=1.0\textwidth]{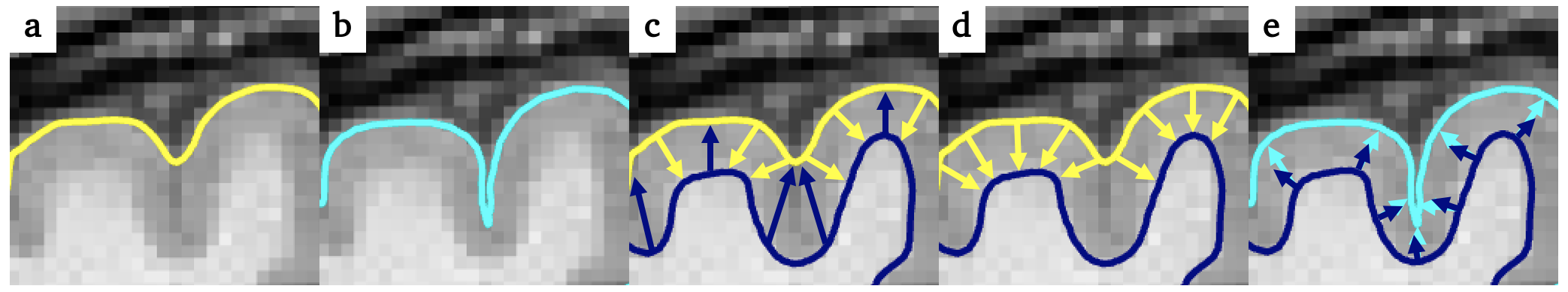}
\caption{T1w brain MRI from the HCP dataset~\cite{van2013hcp}. (a) The cGM segmentation boundary ({\color{Yellow}yellow}); (b) The expected pial surface ({\color{Cyan}cyan}); (c) Bi-Chamfer distance between the cGM segmentation boundary and input white surface ({\color{Blue}blue}). (d) The boundary loss, \emph{i.e.}, single-directional Chamfer distance. (e) The inflation loss between the vertex displacement ({\color{Cyan}cyan}) and the normal vector ({\color{Blue}blue}) of the input white surface.} \label{fig:loss}
\end{figure}

\noindent\textbf{Pial Surface Reconstruction.} Since the cGM segmentations are strongly affected by partial volume effects, it only provides weak supervision especially for low-resolution data. Fig.~\ref{fig:loss}-a shows that the cGM segmentation fails to identify the voxels in the deep cortical sulci. Different from white surface reconstruction, the pial surfaces learned through the bi-Chamfer loss may fail to capture cortical folding and thus cause inaccurate estimation of morphological features.

Inspired by deformation-based approaches~\cite{makropoulos2018dhcp,schuh2017deformable}, we propose a novel weakly supervised loss to tackle partial volume effects on the pial surfaces. The predicted white surface is used as the input mesh $\mathcal{M}_0$. First, we introduce a \textit{boundary loss}:
\begin{equation}\label{eq:boundary-loss}
\mathcal{L}_{\mathrm{boundary}}(\hat{\mathcal{M}}, \mathcal{M}_*)=\frac{2}{|\mathcal{V}_*|}
\sum\nolimits_{v_*\in\mathcal{V}_*}
\min_{\hat{v}\in\hat{\mathcal{V}}}\|v_*-\hat{v}\|^2.
\end{equation}
The boundary loss is equivalent to a single-directional Chamfer loss that computes the shortest distance from the boundary $\mathcal{M}_*$ of the pGT cGM segmentation to the predicted pial surface $\hat{\mathcal{M}}$. As shown in Fig.~\ref{fig:loss}-c,d, the proposed boundary loss will extend the input white surface towards the cGM boundary without affecting the sulcal regions, whereas the bi-Chamfer distance will deform the surface outwards from the deep sulci. Next, an \textit{inflation loss} is defined as:
\begin{equation}\label{eq:inflation-loss}
\mathcal{L}_{\mathrm{inflation}}(\hat{\mathcal{M}},\mathcal{M}_0)=1-\frac{1}{|\hat{\mathcal{V}}|}
\sum\nolimits_{i=1}^{|\hat{\mathcal{V}}|}
\frac{\hat{v}^i-v_0^i}{\|\hat{v}^i-v_0^i\|+\epsilon}\cdot n(v_0^i),
\end{equation}
where $\epsilon$=$10^{-12}$, $v_0^i$ is the $i$-th vertex of the input white surface $\mathcal{M}_0$, and $n(v_0^i)$ is the normal vector with $\|n(v_0^i)\|$=1. The inflation loss measures cosine similarity between the vertex displacement and the normal vectors (Fig.~\ref{fig:loss}-e). This ensures that the inflation of input white surface adheres to its normal direction. The final weakly supervised loss is defined as 
$\mathcal{L}_{\mathrm{weak}}=\mathcal{L}_{\mathrm{boundary}}+w_{\mathrm{inflation}}\mathcal{L}_{\mathrm{inflation}}$.

The weakly supervised loss encourages the input white surface to inflate along the normal until it fits the cGM boundary or two gyri touch each other. Such inflation can alleviate the partial volume effects and allows the pial surfaces to deform into the deep cortical sulci. During the inflation, mesh self-intersections can be prevented by diffeomorphic surface deformations without additional collision detection. Considering the smoothness terms, the final loss for pial surface reconstruction is defined as
$\mathcal{L}_{\mathrm{pial}}=\mathcal{L}_{\mathrm{weak}}+w_{\mathrm{edge}}\mathcal{L}_{\mathrm{edge}}+w_{\mathrm{nc}}\mathcal{L}_{\mathrm{nc}}$.

At the beginning of the training, we have $\hat{\mathcal{M}}=\mathcal{M}_0$ since the input white surface has not been deformed. The gradient of the inflation loss $\mathcal{L}_{\mathrm{inflation}}$ with respect to the $i$-th vertex can be computed as $\nabla_{i}\mathcal{L}_{\mathrm{inflation}}|_{\hat{\mathcal{M}}=\mathcal{M}_0}=-n(v_0^i)/(\epsilon|\hat{\mathcal{V}}|)$. However, this leads to exploding gradients as $\epsilon$ is very small. To address this issue, we pre-train the TA-Net for a few epochs by replacing $\mathcal{L}_{\mathrm{weak}}$ with a MSE loss $\mathcal{L}_{\mathrm{mse}}(\hat{\mathcal{M}}, \mathcal{M}_N)=
\sum_{i=1}^{|\hat{\mathcal{V}}|}\|\hat{v}^i-v_N^i\|^2/|\hat{\mathcal{V}}|$, where $\mathcal{M}_N$ is an inflated white surface with vertices $v_N^i=v_{N-1}^i+\gamma n(v_{N-1}^i)$ for $N$=$1,...,10$ and $\gamma$=0.1. The predicted surface learns to extend outwards during pre-training while avoiding exploding gradients. After pre-training, we resume to use $\mathcal{L}_{\mathrm{weak}}$ as the reconstruction loss.

\section{Experiments}

\noindent\textbf{HCP Young Adult Dataset.} We evaluate CoSeg on the publicly available HCP young adult dataset~\cite{van2013hcp} with 1,113 T1w brain MRI, cortical ribbon segmentations and cortical surfaces extracted by the HCP pipeline~\cite{glasser2013hcp} with quality control. The segmentations are used as the pGT for training, and the pGT cortical surfaces are only used for evaluation. The dataset is split by the ratio of 60/10/30\% for training/validation/testing. All T1w images and segmentations have been affinely aligned to the MNI-152 space. We compute the average segmentations of all training subjects as a template, from which we extract an initial surface with 160k vertices. Both MRI and segmentations are resampled to 1mm/2mm isotropic resolutions and clipped to the sizes of 112$\times$224$\times$176 and 56$\times$112$\times$88 for each brain hemisphere. Without loss of generality, we only consider the left hemisphere in the experiments. We train TA-Nets for 200 epochs by Adam optimizer with learning rate 0.0001. For pial surfaces, we first pre-train the TA-Net using the MSE loss $\mathcal{L}_{\mathrm{mse}}$ for 20 epochs, and then optimize the weakly supervised loss $\mathcal{L}_{\mathrm{weak}}$ for 180 epochs. We use $w_{\mathrm{edge}}$=0.5 and $w_{\mathrm{nc}}$=5.0 for smoothness terms. The integration time and step size are set to $T$=1 and $h$=0.02 with $K$=50 steps. We compare CoSeg with both implicit and explicit learning-based cortical surface reconstruction approaches as shown in Table~\ref{tab:feature}. All experiments are performed on a NVIDIA RTX3080 GPU with 10GB memory. 
\newline

\begin{table}[t] 
\scriptsize
\centering
\begin{minipage}{.55\linewidth}
\centering
\caption{Comparison with existing learning-based approaches in terms of the types of surface representation, required supervision and primary loss functions.}
\label{tab:feature}
\resizebox{1.0\columnwidth}{!}{%
\begin{tabular}{lp{1pt}cp{1pt}cp{1pt}c}
\toprule
Method && Type && pGT && Loss function\\
\midrule
3D U-Net~\cite{ronneberger2015unet} && Implicit && Seg && Cross Entropy\\
DeepCSR~\cite{cruz2021deepcsr} && Implicit && SDF && L1 Loss\\
Vox2Cortex~\cite{bongratz2022vox2cortex} && Explicit && Surface && Bi-Chamfer\\
CortexODE~\cite{ma2022cortexode} && Explicit && Surface && Bi-Chamfer\\
CFPP~\cite{santa2022cfpp} && Explicit && Surface && Bi-Chamfer\\
CoTAN~\cite{ma2023cotan} && Explicit && Surface && Bi-Chamfer\\
CoSeg (Ours) && Explicit  && Seg && Weakly Supervised\\
\bottomrule
\end{tabular}
}
\end{minipage}
\hfill
\begin{minipage}{.42\linewidth}
\centering
\caption{Comparison with implicit approaches on the HCP dataset at 1mm/2mm resolutions. The ASSD (mm) and HD90 (mm) are reported.}
\label{tab:implicit}
\resizebox{1.0\columnwidth}{!}{%
\begin{tabular}{clp{1pt}cccc}
\toprule
& && \multicolumn{2}{c}{White} & \multicolumn{2}{c}{Pial} \\
& Method &&
ASSD$\downarrow$ & HD90$\downarrow$ &
ASSD$\downarrow$ & HD90$\downarrow$ \\
\midrule
\multirow{3}{*}{\rotatebox[origin=c]{90}{1mm}}~
& 3D U-Net &&
0.180 & \textbf{0.373} &
0.432 & 1.257 \\
& DeepCSR &&
0.213 & 0.468 &
0.436 & 1.456\\
& CoSeg &&
\textbf{0.177} & 0.389 &
\textbf{0.257} & \textbf{0.553} \\
\cmidrule{1-2}\cmidrule(lr){4-5}\cmidrule(lr){6-7} 
\multirow{3}{*}{\rotatebox[origin=c]{90}{2mm}}~
& 3D U-Net &&
0.385 & 0.831 &
0.833 & 3.125 \\
& DeepCSR &&
0.404 & 1.019 &
0.904 & 4.007 \\
& CoSeg &&
\textbf{0.330} & \textbf{0.756} &
\textbf{0.361} & \textbf{0.767}\\
\bottomrule
\end{tabular}
}
\end{minipage}
\end{table}

\begin{figure}[t]
\centering
\includegraphics[width=0.98\textwidth]{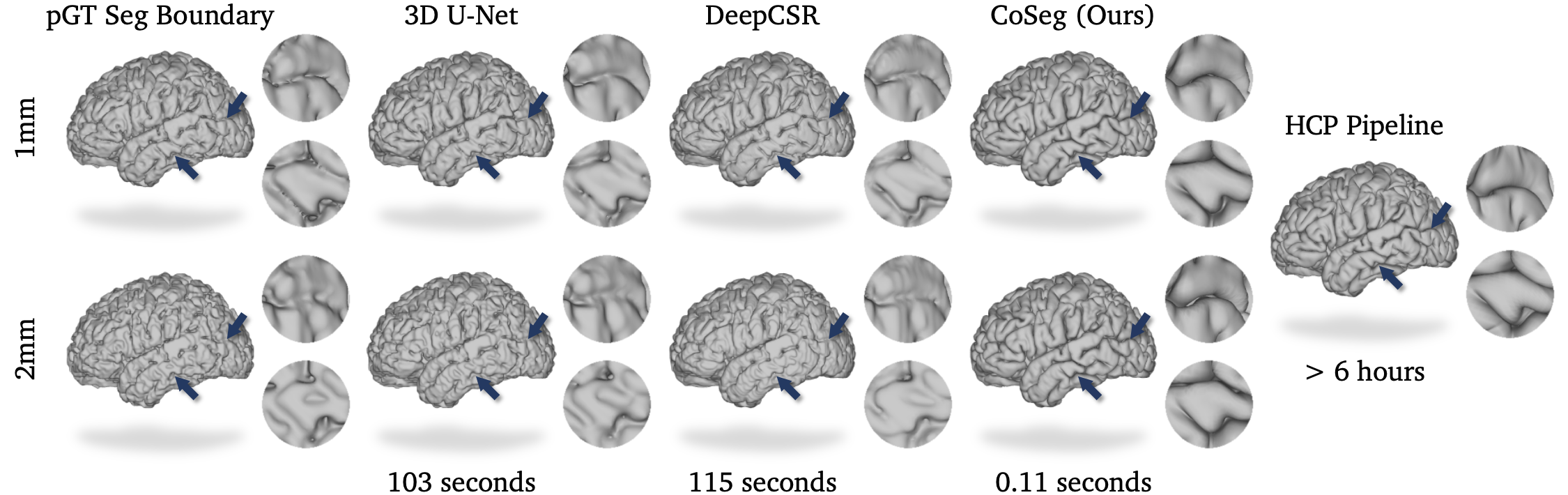}
\caption{Pial surfaces predicted by implicit approaches on the HCP dataset. The runtime required for each brain hemisphere (including topology correction) is reported.} \label{fig:implicit}
\end{figure}

\noindent\textit{Comparison with Implicit Approaches:} We compare CoSeg with implicit learning-based approaches for 1mm and 2mm resolutions. For CoSeg, we train two TA-Nets for white and pial surfaces supervised by the pGT segmentation boundary. The weights $w_{\mathrm{inflation}}$ are set to 2.0/5.0 for 1mm/2mm resolutions. Since the main backbone of the TA-Net is a 3D U-Net~\cite{ronneberger2015unet}, we train U-Nets to learn segmentations for fair comparison. We also compare to DeepCSR~\cite{cruz2021deepcsr}, of which the pGT SDFs are created by the distance transform~\cite{breu1995distance} of pGT segmentations. The surface meshes ($|V|$$\approx$150k) are extracted by MC followed by topology correction~\cite{bazin2005topology} and Taubin smoothing~\cite{taubin1995curve}. The performance is evaluated by the average symmetric surface distance (ASSD) and 90th-percentile Hausdorff distance (HD90)~\cite{bongratz2022vox2cortex,cruz2021deepcsr,ma2022cortexode} between the predicted and pGT cortical surfaces.

Table~\ref{tab:implicit} reports that CoSeg achieves superior geometric accuracy except for white surfaces at 1mm. Fig.~\ref{fig:implicit} shows that CoSeg is able to deform the pial surfaces into deep sulci, while the implicit approaches fail to capture the cortical sulci due to intrinsic partial volume effects of volumetric data, leading to large geometric errors for pial surfaces. CoSeg only requires 0.11s of runtime for each brain hemisphere. This is orders of magnitude faster than the HCP pipeline, U-Net and DeepCSR. CoSeg only produces a negligible number of self-intersecting faces, \emph{i.e.}, 0.06/3.65 out of 330k faces on average for white/pial surface.
\newline

\begin{table}[t]
\scriptsize
\centering
\caption{Comparison with explicit learning-based approaches and ablation studies of CoSeg on the HCP dataset at 1mm/2mm resolutions. The ASSD (mm), HD90 (mm), errors of cortical thickness $e_{\mathrm{thick}}$ (mm) and sulcal depth $e_{\mathrm{sulc}}$ (mm) are reported. The best results are in bold and the second best results are underlined.}
\label{tab:explicit}
\begin{tabular}{clp{1pt}ccccp{1pt}cccc}
\toprule
& && \multicolumn{4}{c}{Resolution (1mm)} & &\multicolumn{4}{c}{Resolution (2mm)} \\
& Method && ASSD$\downarrow$ & HD90$\downarrow$ &
$e_{\mathrm{thick}}\downarrow$ & $e_{\mathrm{sulc}}\downarrow$ && ASSD$\downarrow$ & HD90$\downarrow$ &
$e_{\mathrm{thick}}\downarrow$ & $e_{\mathrm{sulc}}\downarrow$ \\
\midrule
& Vox2Cortex~\cite{bongratz2022vox2cortex} && 
0.292 & 0.641 & 0.511 & 1.238 &&
0.542 & 2.026 & 0.731 & 1.432 \\
& CortexODE~\cite{ma2022cortexode} & &
0.352 & 0.796 & 0.448 & 0.672 &&
0.610 & 1.787 & 0.742 & 1.303 \\
& CFPP~\cite{santa2022cfpp} & &
0.274 & 0.671 & 0.496 & 1.417 &&
0.578 & 2.385 & 0.872 & 1.647 \\
& CoTAN~\cite{ma2023cotan} & &
0.293 & 0.769 & 0.534 & 1.438 &&
0.538 & 1.966 & 0.781 & 1.781 \\
\cmidrule(lr){1-2}\cmidrule(lr){4-7}\cmidrule(lr){9-12}
& CoSeg ($w_{\mathrm{inflate}}$=0.0) & &
\underline{0.253} & \underline{0.564} & 0.379 & 1.389 &&
0.383 & 0.980 & 0.551 & 1.580  \\
& CoSeg ($w_{\mathrm{inflate}}$=2.0) & &
\textbf{0.252} & \textbf{0.539} & \textbf{0.298} & \underline{0.530} &&
\textbf{0.347} & \textbf{0.769} & 0.424 & \underline{0.686} \\
& CoSeg ($w_{\mathrm{inflate}}$=5.0) & &
0.273 & 0.627 & \underline{0.320} & \textbf{0.435} &&
\underline{0.351} & \underline{0.777} & \textbf{0.386} & \textbf{0.397}  \\
& CoSeg ($w_{\mathrm{inflate}}$=10.0) & &
0.298 & 0.736 & 0.362 & 1.120 &&
0.370 & 0.881 & \underline{0.421} & 0.797  \\
\bottomrule
\end{tabular}
\end{table}

\begin{figure}[t]
\centering
\includegraphics[width=0.98\textwidth]{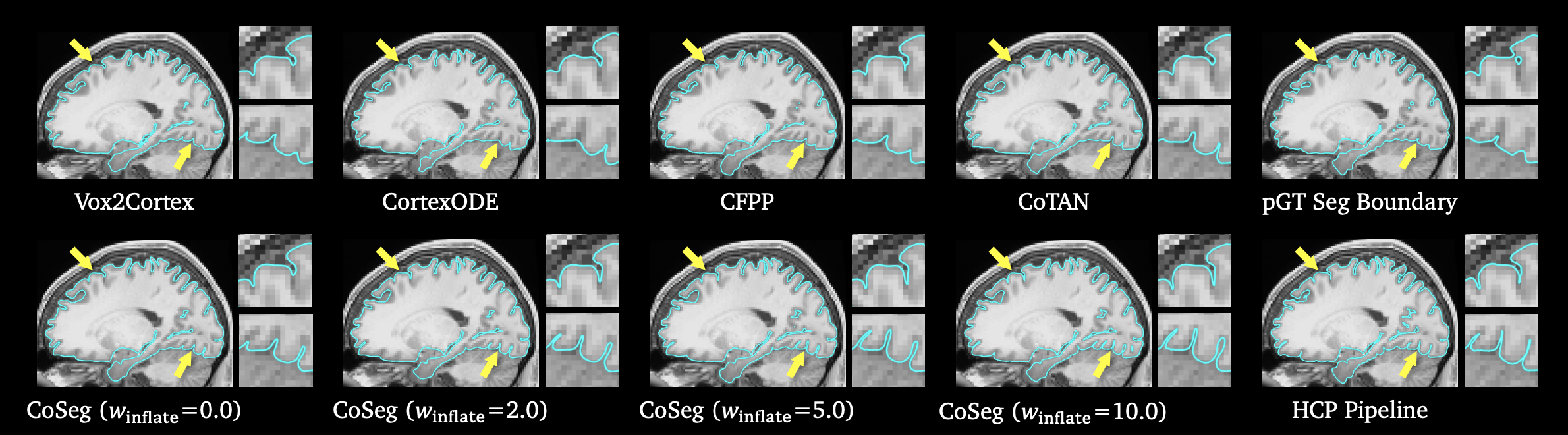}
\caption{Pial surfaces (2mm) predicted by explicit approaches on the HCP dataset.} \label{fig:explicit}
\end{figure}

\noindent\textit{Comparison with Explicit Approaches:}
We compare CoSeg with explicit learning-based approaches including Vox2Cortex~\cite{bongratz2022vox2cortex}, CortexODE~\cite{ma2022cortexode}, CorticalFlow++ (CFPP)~\cite{santa2022cfpp} and CoTAN~\cite{ma2023cotan}. For CoTAN we use the same architecture as the TA-Net for adult data. To show the advantage of proposed weakly supervised loss, we train all approaches to learn pial surfaces from the pGT segmentation boundary at 1mm/2mm resolutions. The pGT white surfaces are used as the input for training, such that the predicted and pGT pial surfaces have the same mesh connectivity. This allows vertex-to-vertex comparison of cortical morphological features~\cite{glasser2013hcp}. We compute the average L1 errors of the cortical thickness and the sulcal depth of midthickness surfaces over all vertices between the predicted and pGT cortical surfaces. The geometric accuracy is measured by ASSD and HD90. We also conduct ablation studies to explore the influence of different weights $w_{\mathrm{inflation}}$ for the inflation loss $\mathcal{L}_{\mathrm{inflation}}$ on CoSeg.

As shown in Table~\ref{tab:explicit} and Fig.~\ref{fig:explicit}, CoSeg achieves superior geometric and morphological accuracy. Existing explicit approaches~\cite{bongratz2022vox2cortex,ma2023cotan,ma2022cortexode,santa2022cfpp}, which use the bi-Chamfer loss (see Table~\ref{tab:feature}), overfit to the pGT segmentation boundary and thus fail to fit the deep cortical sulci especially for low resolution samples. This results in inaccurate morphological features, \emph{i.e.}, greater cortical thickness and shallower sulcal depth (see Appendix). Our ablation study shows that the inflation loss effectively reduces morphological errors. However, excessive weight for the inflation loss also affects the performance, since the vertex displacement of the pial surface may not strictly follow the normal direction of the white surface.
\newline

\begin{figure}[t]
\centering
\includegraphics[width=1.0\textwidth]{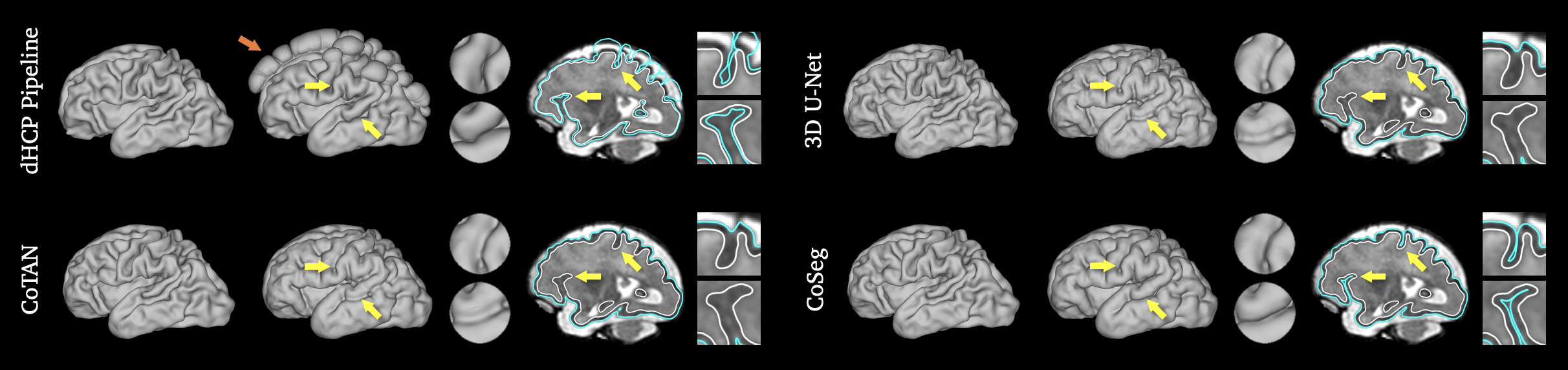}
\caption{Cortical surfaces of a 32.4-week subject in the dHCP fetal dataset.} \label{fig:fetal}
\end{figure}

\noindent\textbf{dHCP Fetal Dataset.} 
We evaluate CoSeg on the dHCP fetal dataset~\cite{price2019fetal} with 241 T2w fetal brain MRI scanned at the ages of 22--38 weeks. The dataset is split by the ratio of 60/10/30\%. The voxel size of the T2w images is 0.5mm$^3$. However, this is a still relatively low resolution as the fetal brains are very small. The pGT tissue segmentations are created by the BOUNTI pipeline~\cite{uus2023bounti}, a fully validated learning-based fetal brain segmentation approach. The pGT cortical ribbon segmentations are created by merging the tissue labels. Both T2w images and segmentations are affinely aligned to a 36-week fetal brain MRI atlas~\cite{uus2023atlas} and clipped to the size of 112$\times$224$\times$176 for each hemisphere. An initial surface is generated based on the segmentation atlas~\cite{uus2023atlas}. For CoSeg we set $w_{\mathrm{inflation}}$=5.0.

We compare CoSeg with the dHCP neonatal pipeline~\cite{makropoulos2018dhcp}, CoTAN~\cite{ma2023cotan} and 3D U-Net supervised by pGT segmentations. Since the Dice score cannot reflect the geometric accuracy, we provide qualitative comparisons in Fig.~\ref{fig:fetal}. It shows that CoSeg predicts high-quality surfaces, while the dHCP neonatal pipeline fails to generalize well on fetal subjects. The pial surfaces predicted by U-Net and CoTAN are severely affected by the partial volume effects of the cGM.

\section{Conclusion}
In this work, we propose CoSeg for diffeomorphic cortical surface reconstruction weakly supervised by cortical ribbon segmentations. A novel weakly supervised loss is designed to address partial volume effects and deform the pial surfaces into deep cortical sulci. CoSeg shows remarkable results on both HCP young adult and dHCP fetal dataset without the need for pGT cortical surfaces. One limitation is that CoSeg guides vertex displacement strictly along the normal of the input white surface, with potential to influence anatomical fidelity of pial surfaces. The efficacy of CoSeg is also affected by the quality of pGT segmentations. Looking ahead, we plan to explore semi-supervised learning, utilizing both labeled segmentations and unlabeled MRI intensity to refine the performance.
\newline

\noindent\textbf{Acknowledgments:}
This work is partially supported by the President’s PhD Scholarship at Imperial College London. Support was also received from the ERC project MIA-NORMAL 101083647 and ERC project Deep4MI 884622. Further support was received by the State of Bavaria (HTA). HPC resources provided by the Erlangen National High Performance Computing Center (NHR@FAU) of the Friedrich-Alexander-Universität Erlangen-Nürnberg (FAU) under the NHR project b180dc. NHR@FAU hardware is partially funded by the German Research Foundation (DFG) – 440719683. The dHCP fetal data were provided by the developing Human Connectome Project, KCL-Imperial-Oxford Consortium funded by the ERC under the European Union Seventh Framework Programme (FP/2007-2013) / ERC Grant Agreement no. [319456].

%
%
%
\bibliographystyle{splncs04}
\bibliography{paper1517}
\end{document}